\documentclass[conference,10pt]{IEEEtran}
\IEEEoverridecommandlockouts
\usepackage{cite}
\usepackage{amsmath,amssymb,amsfonts}
\usepackage{algorithmic}
\usepackage{graphicx}
\usepackage{textcomp}
\usepackage{xcolor}
\usepackage{hyperref}
\usepackage[ruled,vlined,linesnumbered]{algorithm2e}
\def\BibTeX{{\rm B\kern-.05em{\sc i\kern-.025em b}\kern-.08em
    T\kern-.1667em\lower.7ex\hbox{E}\kern-.125emX}}
\begin{document}

\title{An Incentive Based Approach for COVID-19 using Blockchain Technology
}

\author{\IEEEauthorblockN{Manoj MK\IEEEauthorrefmark{1},
		Gautam Srivastava\IEEEauthorrefmark{2}, Siva Rama Krishnan Somayaji\IEEEauthorrefmark{1},\\ Thippa Reddy Gadekallu\IEEEauthorrefmark{1}, 
		Praveen Kumar Reddy Maddikunta\IEEEauthorrefmark{1}, and Sweta Bhattacharya\IEEEauthorrefmark{1}}
	\IEEEauthorblockA{
		\IEEEauthorrefmark{1}Vellore Institute of Technology, Vellore, Tamil Nadu, India  \\
		\{mk.manoj2015,siva.s,thippareddy.g,praveenkumarreddy,sweta.b\}@vit.ac.in\\
		\IEEEauthorrefmark{2}Department of Mathematics and Computer Science, Brandon University, Manitoba, Canada; srivastavag@brandonu.ca \\
		}}

%

\maketitle

\begin{abstract}
The current situation of COVID-19 demands novel solutions to boost healthcare services and economic growth. A full-fledged solution that can help the government and people retain their normal lifestyle and improve the economy is crucial.  By bringing into the picture a unique incentive-based approach, the strain of government and the people can be greatly reduced. By providing incentives for actions such as voluntary testing, isolation, etc., the government can better plan strategies for fighting the situation while people in need can benefit from the incentive offered. This idea of combining strength to battle against the virus can bring out newer possibilities that can give an upper hand in this war. As the unpredictable future develops, sharing and maintaining COVID related data of every user could be the needed trigger to kick start the economy and blockchain paves the way for this solution with decentralization and immutability of data.
\end{abstract}

\begin{IEEEkeywords}
COVID-19, coronavirus, Blockchain, Incentive based token, Travel history, tamper proof data, colour bands.
\end{IEEEkeywords}

\section{Introduction}

The novel coronavirus or COVID-19 is a global pandemic respiratory disease caused by SARS-CoV2. The World Health Organization(WHO) declared the outbreak a pandemic on March 11th 2020 \cite{WHO}. Some of the preventive measures include repeated hand washing, usage of face mask and social distancing \cite{PublicHealth}. The virus spreads from person to person mainly through contact with infected secretions. The ongoing pandemic has caused serious social and economic disruptions globally. The shortage of medical supplies, infrastructure and daily necessities is just the tip of the problems faced. The global recession, lockdown and closing of schools or universities, etc. had a major setback on mental and physical health of the people worldwide. The stress and overwork of frontline workers during these tough times are showcased severely by media. \cite{yuen2020psychological}
The key to winning this battle is through effective isolation of COVID positive patients and their close contacts. Until a vaccine is ready, there are no other means to stop the virus spreading. During these severe economic crises, a vast majority of people are laid off and this in conjunction with the closing of shops, factories, etc. It is safe to assume that the economic cash flow is dwindling. This situation could be a battle as severe as the pandemic itself. Every country has taken varied approaches to combat these problems. 

The use of information and communication technology plays a significant role in helping healthcare workers perform the diagnosis faster\cite{gadekallu2020deep,gadekallu2020early}. The deployment of body wearable sensors, telehealth and AI-chatbots, can help in remote diagnosis of the patients \cite{healthcareitnews,abou2020ditrust}. Various AI applications play a crucial role in predicting virus effects by analyzing previous or historical data \cite{vaishya2020artificial}. In countries like Afganistan, where health care services have resource constraints, the deployment of Internet of things (IoT) along with telemedicine, was considered \cite{azizy2020not}. After the lockdown, a few organizations have begun restoring their services by following strict guidelines such as sanitization, social distancing and thermal scanning before they allow customers to enter the premises. In a few places, the researchers have implemented a smart helmet with Mounted Thermal Imaging System embedded with IoT for obtaining real-time data. This is one of the solutions to curtail the virus spread \cite{hakak2020have}.

The principal objective of this paper is to help bridge the gap of the economy discrepancy while curtailing the virus spread.  Since the virus can only spread from person to person, by backtracking all contacts of the origin, an effective strategy of isolation can be implemented. A unique incentive-based approach is provided to channel isolation while also helping the people in need during these tough times. The use of blockchain\cite{bera2020blockchain,aggarwal2019blockchain,hakak2020recent} in this work provides a means of decentralization of information that is severely required to propagate effective battle plans to not only scrub the virus at current times but also to think about the steps to sail through the uncertain future effectively. 

The unique contributions of the work are given below:
\begin{enumerate}
\item A unique incentive based approach for effective cooperation and battle against the pandemic to provide both the government and its people a win-win situation
\item A blockchain based solution to prevent information tampering such as the covid test results
\end{enumerate}

The organization of the paper is as follows.  Section 2 presents the Literature review. Section 3 presents preliminary information on the technology used and describes the proposed methodology.  Section 4 highlights the results of experimentation, and Section 5 discusses the conclusions and points direction of future work.

\section{Literature Survey}
With the advent of development in new technologies such as 5G, Internet of medical things (IOMT), edge computing, artificial intelligence (AI), blockchain, etc., solving real-world problems become more natural and efficient. In most of the scenarios presented, the AI and IOMT work together on the blockchain framework. From the end of 2019, the world is facing a massive health crisis due to the rapid spread of the coronavirus. The authors in \cite{ting2020digital} analyze the integration of new technologies to aid in the diagnosis of COVID-19. The authors also infer from the survey that the blockchain technology can help in accelerating the diagnostic process.

The world has come to a standstill due to the epidemic spread of coronavirus, and it has affected all the sectors drastically. The current health care system may not be effective is quick diagnosis and treatment on a larger scale. The authors in \cite{mashamba2020blockchain} propose a cost-effective framework for integrating blockchain technology and AI for testing and tracking COVID-19 patients. This can be an effective solution during this crucial resource constraint situation.

The accumulation of funds to help the COVID-19 is vital. These funds need to be adequately managed to dissipate the resources effectively. The integration of blockchain with IPFS and AI \cite{johnstone2020viral} can be an efficient solution for this challenge. The blockchain can be used for a secure data store of the records and the related parameters inputted by the AI techniques.

The epidemic virus called 2019-nCoV was reported by the Chinese Govt. in the first week of January 2020. The WHO has listed various strategies to be followed for detecting and preventing the spread of this epidemic. The technologists and researchers worldwide are working on effective strategies to be implemented in the health care sector for better and faster diagnoses. The authors in \cite{khatoon2020use,chang2020can} propose blockchain as a framework to store, track, and securely share medical data. The authors also use consensus mechanisms such as a smart contract to aid the secured data sharing process.

The coronavirus epidemic has affected all the sectors across the globe. There is a dire need for a fast and accurate system for diagnosing the several patients affected by COVID-19. The authors analyze various technologies \cite{pham2020artificial}, which can be used to aid the diagnostic process. Most of the researchers talk about using image processing and AI to analyze the patients' records. Big data analytics also plays a vital role in analyzing the virus's symptoms for creating the drug. Many researchers across the globe are working together and sharing data over blockchain using a consensus method.  

Various policies and strategies are being formed by several countries to combat the crisis during the lockdown phase of COVID-19. This lockdown has affected the global economy severely, and all the countries joined together to find the solution in the current situation. The authors in \cite{allam2020coronavirus} highlight the use of smart city network for secure data sharing across different countries to formulate newer policies which help in increasing the overall economic development. The authors also suggest using blockchain technology while sharing the data securely and maintaining the privacy of the stored data.

Blockchain technology revolutionized the health care sector by providing a secure means of transaction processing and data access ubiquitously. Every bit of detail of electronic health records is accessible through blockchain, which aids easier and a secured access \cite{deepa2020survey} controlled by the patient.  

Researchers and medical practitioners across the globe are trying to find a suitable vaccine for the coronavirus. In this regard, the upcoming challenge would be to distribute this vaccine to the entire world. The authors in \cite{ramirez2020blockchain} propose blockchain based supply chain monitoring system to help distribute the vaccine. This recommended system aids in delivering the vaccine in 3 stages; sent, storage and application, where the blockchain is used to monitor each step of the vaccine supply chain.

There is a massive effect of COVID-19 on the economic conditions of developing nations. Most of the production units are in China, and the financial crisis also affects these big companies. The existing supply chain methods are failing to address the material shortage, which might continue post-pandemic. The authors in \cite{priyankablockchain} suggest after thorough analysis that the blockchain technology would help is improvising the supply chain process. The authors also indicate that smart contracts can eradicate the delays that occurres due to paperwork.

The information posted on social media has a significant influence on the mindset of people. Unfortunately, information shared on any social media platform is rarely accurate and requires verification with different sources. There is an urgent need to capture the real information posted on social media, and a proper method needs to be established on propagating such information. The authors in \cite{li2020characterizing} use Weibo data and natural language processing methods to segregate the COVID-19 information into situational details. The authors also analyze specific features in estimating the social media post, which might provide useful information to the public.

During this pandemic situation, all the countries are using innovative methods to spread awareness of COVID-19. For example, the Australian government came up with a tracking smartphone app \cite{nabben2020trustless}, which is used to deliver the current situation and information to the people. This app gives significant privacy issues and provides an unconventional policy addressing the privacy protection and trust-less IT infrastructure to develop Australia's digital-political comeback to COVID-19.

COVID-19 has challenged the current health systems and also the global economy. Most of the countries are encouraging scientists, medical practitioners, and academicians to research upon the solution for combating the current crisis. The authors in \cite{chesbrough2020recover} suggest that open innovation would help in finding the solutions faster. Open innovation provides a platform for any user from any domain irrespective of this background to get involved in delivering solutions that might improve the crisis.   

There is a dire need for a stable data management and information sharing method, which can aid in speeding up the diagnosis or the discovery of the vaccine for COVID-19. The existing technologies such as database or cloud has main challenges such as security\cite{iwendi2020keysplitwatermark} and a single point of failure of a centralized system. The authors in \cite{azim2020blockchain} propose using blockchain technology for data management and storage, which would safeguard patients' data stored and would provide robust data management to track the affected patients.

The vaccine for COVID-19 is still being researched and might be released after thorough testing, which is months away. Meanwhile, most of the countries have started resuming regular operation but with maintaining social distancing and sanitizer usage. The authors in \cite{choudhury2020covidchain} implement a novel solution using an app in the smartphones to track the location history and prove that it can be used as an e-pass to confirm that the individual is not in contact with a person infected by the virus. This data is stored in the disturbed ledger of blockchain, which is securely available to any official. In a similar work of tracking the users' location, the authors in \cite{xu2020beeptrace} implement BeepTrace, which is a method of tracking the contact of the individual. The users' credentials and the location information is stored securely in the blockchain.

\section{Proposed System}

A blockchain based system is proposed to keep track of every individual’s COVID data in the country while also maintaining a personal color scheme that can indicate whether an individual is tested positive, is suspected of having the virus or is healthy.
One of the proposal's main ideas is to keep track of an individual's passport number to identify the travel history to segregate and isolate potential cases. This is done so that the virus may be contained at its initial stage without much spreading. Every individual is provided with a unique identifier that may be different according to the country, such as the case of Social Security Number, etc. This is used as an identification with combined reference to the individual's passport number to identify an individual uniquely. This is proposed since the proper identification of an individual is crucial for preventing the virus from spreading. Misidentification of an individual can cause isolation of a wrong individual while not helping prevent the infection from spreading.

The key idea is to verify an individual's blockchain data to provide a means of the tamper-proof record of a person's condition. This is a dire need of the hour since there may be defaulters who may not realize the situation's seriousness and can negatively impact it.
Three colour bands are proposed, such as Red for Positive, Amber for suspicion, or awaiting test result and Green for a healthy individual. The primary spread of the virus is through a person who has already contracted the disease. Thus, by identifying the root origin and segregating the individuals to isolation, the virus's spread can be stopped to a great extent. By searching the travel history of each individual, the probability that people may be affected can be determined before the wider spread. Since the data of the virus spread is relatively known, by analyzing an individual's travel history through the passport number in search with the current situation of the virus at the airports visited by the individual, a clear case of suspected patients may be drawn. To make the search more effective, the spread of virus 14 days before and 14 days after the travel of the individual to the airport may be seen to place an individual under suspicion if the area surrounding the airport in the travel history has had cases. The most efficient solution would be to determine and track the individual's movement completely and every location visited but this data is not readily available and takes a long time to gather. Hence checking the airports visited can start as a small and significant step towards the containment of the virus. Once an individual is placed under suspicion or tested positive, a detailed overview may be obtained from the patient to contain the places the patient had visited.

The usage of blockchain ensures that everything is made digital and now has an easier reference of access to anyone, which helps in identifying more potential cases. Every location must check an individual's data block to obtain their band status before assuring services. This minor improvement can mean everyday activities can resume without hindrance or fear of the virus. The proposed system also contains additional information as a text-based data entry, which can enable flexibility to the deployed blockchain. The need for change in data columns stored on to the blockchain may wary at a later point of time and by deploying a smart contract protocol without any flexibility which means that forking or restarting the chain might be the only solution to implement changes. Hence this addition can effectively improve the adaptability of the blockchain to the unprecedented future needs.

The government has a significant role to play at dire times, such as lockdown wherein every individual is void of cash flow, and many people may suffer from its consequences. As such, the responsibility and the strain on the government officials are too high. There may be cases where keeping track of every individual may not be possible by traditional methods and that obtaining data of every individual currently residing in the country or state may not be possible. This may happen significantly to the lower classes of the society like daily wage workers, homeless, etc., who cannot sustain themselves during a crucial period of lockdown. As a government, to connect with and help everyone may not be possible at times. 

The proposed solution, an incentive-based approach, can combat this issue with very significant benefits. The aim is to provide individuals with Incentive Tokens from the government directly to the individual. A token may be offered to the individual for voluntary testing or agreeing to self-quarantine properly. This enables the government to keep track of the people who willingly cooperate. This means that any individual that may have been missed out can also now be identified and kept track on. The incentive tokens can act as a means to leverage people in abiding by the government created rules and the compensation for an individual who follows the rules could be a token. The token may be utilized by every country to amend and modify the range of rules and its benefits offered. 

The token may act as a direct monetary benefit or a substantial benefit for people who may be in dire need of benefits from their government. There may be situations where the government may be overwhelmed by the situation, and any benefits planned for people may not reach them in their time of need. Such a situation is worse than the virus spread, and the fate such people have to endure may be cruel. 

To combat these potential problems, the incentive token may pave the way while not only having the benefit of co-operative people but also providing direct benefits instead of it reaching various intermediaries before reaching the people. The tokens may be improved by the government according to the needs, but some examples could be, providing monetary benefit, helping provide necessities as a package for exchange of a token, reduction in tax, electricity bill, water bill, reduction of rent, fare reduction for immigrant workers, etc.

The approach can be a significant winner since not only are people able to obtain these benefits in their needy situation; the government can also leverage the incentives to make a broader impact on safety and security by containing the virus spread at a much larger pace. Consider an example, an incentive token may be provided to all individuals who voluntarily take up COVID testing at a potential hotspot location. This could
	benefit not only the people but also the government in helping to assess the situation of the area more accurately while helping to deploy battle plans much efficiently.

\begin{algorithm}[]
	\textbf{Step 1:} Identify $P_v$ where $P_v$ = volunteer\\
	\textbf{Step 2:} \\
	\eIf{(uid)$P_v$ exist}{
		(uid)$P_v$ =unique id assigned to $P_v$\\
		continue;\\
	}
	{
		assign uid to $P_v$
	}
	\textbf{Step 3:} Government Authority increments incentive token count (I) $P_v$\\
	\textbf{Step 4:} Update (I) $P_v$  to the blockchain\\
	\textbf{Step 5:} $P_v$ can redeem (I) $P_v$   based on Government regulations\\
	\caption{Working of incentive token}
	\label{a1}
\end{algorithm}

\section{Results and Discussion}

\begin{figure}[h!]
    \centering
    \frame {\includegraphics[width=0.6\linewidth,height=5cm]{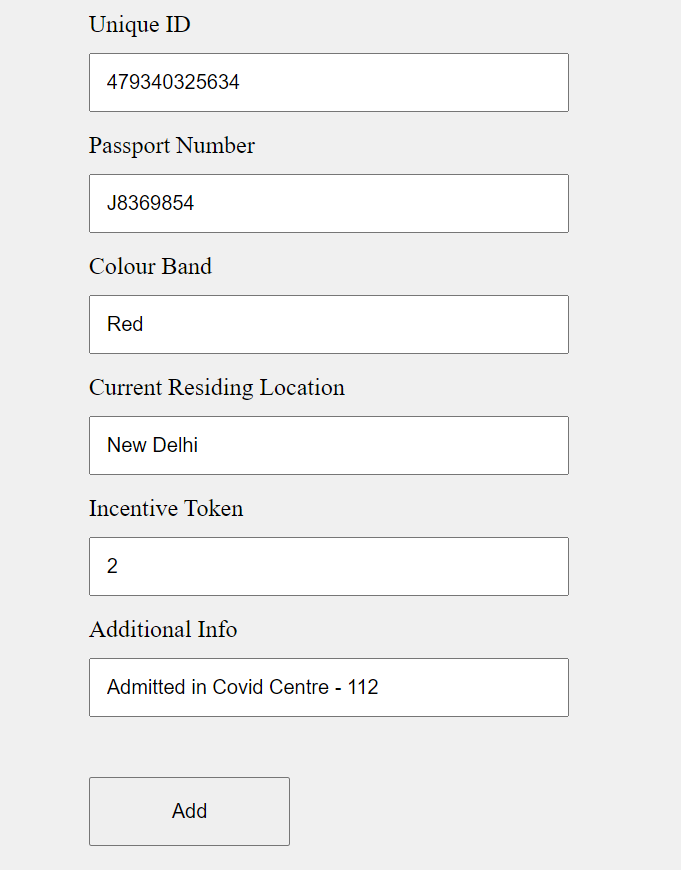} }
    \caption{Add User UI.}
    \label{Fig:2}
\end{figure}

\begin{figure}[h!]
    \centering
    \frame{\includegraphics[width=0.6\linewidth]{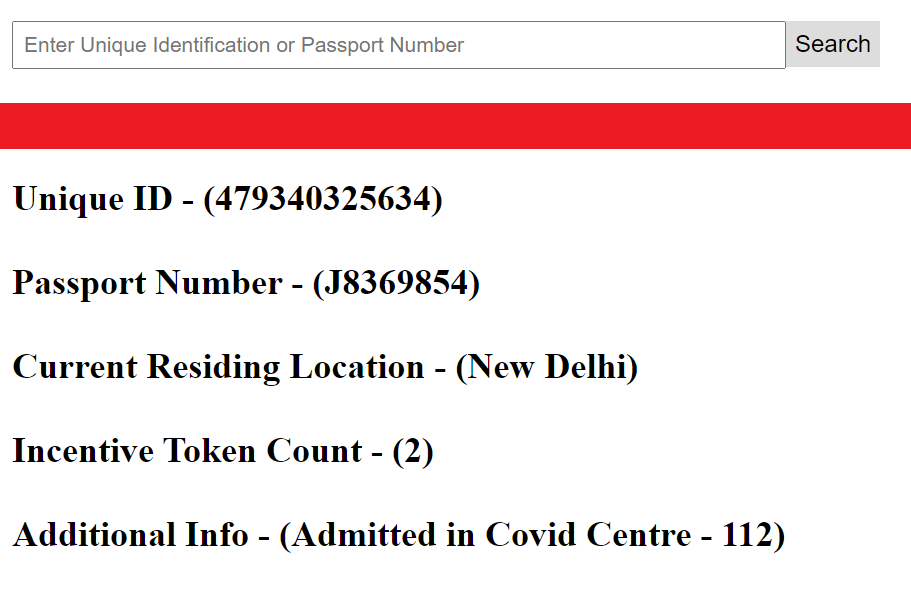}}
    \caption{User Search UI.}
    \label{Fig:3}
\end{figure}

\begin{figure}[h!]
    \centering
\frame{\includegraphics[width=0.6\linewidth]{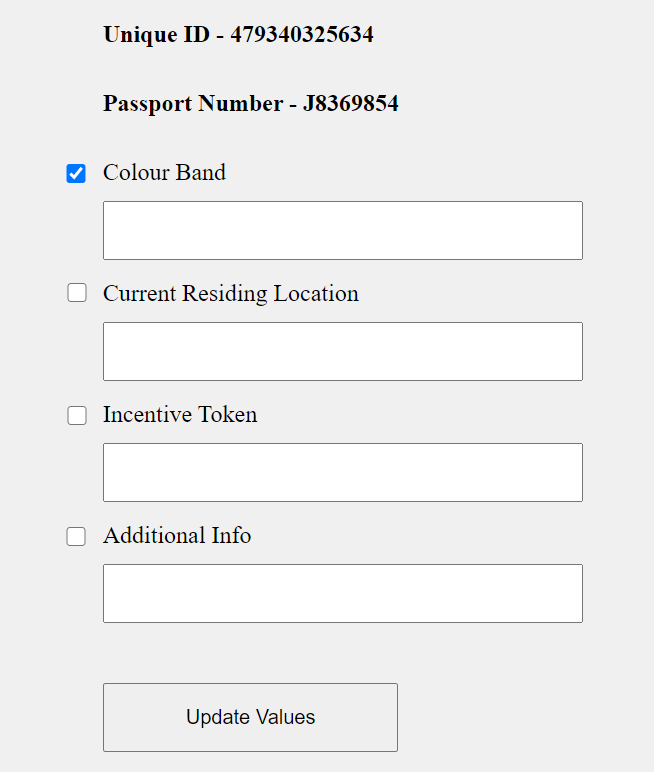}}
    \caption{Updation UI.}
    \label{Fig:4}
\end{figure}

\begin{figure}[h!]
    \centering
    \frame{\includegraphics[width=1.0\linewidth,height=8cm]{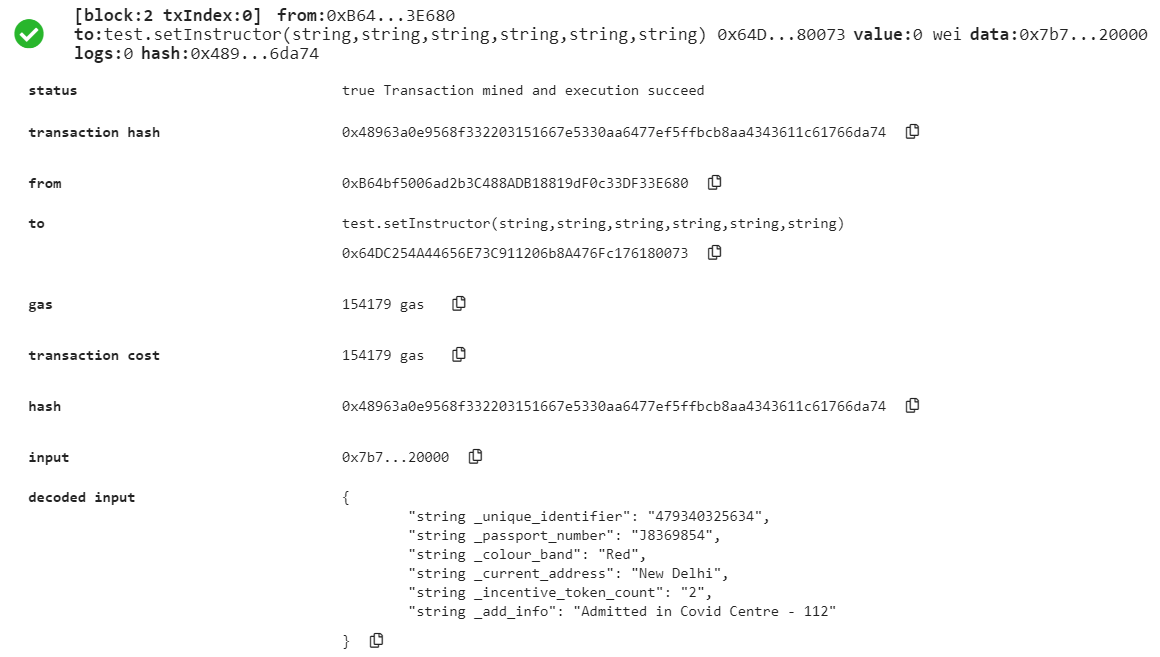}}
    \caption{Sample Block.}
    \label{Fig:5}
\end{figure}

The proposed system can benefit by not only keeping track of individuals in the country but can also act as a means of direct benefit to the individual involved without the need for third-party interference. Users may be added with the help of UI proposed in Fig \ref{Fig:2}. Any user may be added to the system by assigning a unique ID for that user. This can help combat the identification problem of users who may not have registered with the government or were stranded by the current situation. The passport number may not be available for all and hence may be left blank if necessary. Any individual can avail the benefits of the incentive tokens without any repercussions. This token value may be updated along with the colour band, additional info section, and current location through the Update UI, as shown in Fig \ref{Fig:3}. Data of any user may be viewed using the User Side UI by searching for the Unique ID or passport number, as shown in Fig \ref{Fig:4}. This feature may be used before providing services such as restaurant dine-in and can act as a protective barrier in eliminating the virus spread while still paving the way to day to day activities. A sample block created for a user is depicted in Fig \ref{Fig:5}.

The need for keeping track of every individual in the country is enormous and is the need of the hour. By also providing incentive benefits, this method can attract more individuals to join the battle plan of government willingly as it is a win-win situation for both the involved parties. The government now has a record of all the people and can plan to fight the pandemic more efficiently while the users can claim direct benefits from the government. During difficult times such as lockdown, the government's stimulus may be required to sustain a lot of people who may otherwise be unable to fight the situation on their own. Being advantageous to both parties, the proposed system is unique and effective in gathering people willing to fight the pandemic.   

\subsection{Significance of the work}

The proposed system aims to eradicate the virus's entry and spread from a foreign region. By isolating the entry of the virus, the infection can be nipped in its bud. The use of passport to identify and segregate potentially suspected people is the first phase of the battle plan to halt the virus's entry. This way, the potential losses due to the entry of people into a region without proper screening may be significantly reduced. 

In the current era of information, misinformation may be the greatest enemy. To combat this problem, a blockchain-based system is proposed as the core solution. This provides a transparent and immutable copy of the data, freely accessible to anyone. This provides the basis for tracking and isolation of future cases and helps visualize the impact on various regions, which can be used to effectively formulate battle plans such as restriction of services or lockdown to stop the virus in areas of hotspots.

During such drastic measures, the key is to reduce contact between people, which in turn stops the virus spread. But any service, manufacturing, health, or other activities all require interactions of various people. While stopping such activities for a short time may be the most effective plan, its severity can easily be felt by any person. The livelihood of many may be at stake through such a move. The proposed system combats such issues and rewards people for co-operation. By co-operating with the government laid rules, one can earn an incentive token. These tokens, as advised by government may carry a wide range of benefits from monetary benefits to tax reduction, etc. The greatest strength of the token is its flexibility in adapting to any government rules. Not all locations require the same benefits as others, and to cater to the needs of the widespread, the distribution and redeeming of tokens may be varied while still maintaining the general idea of uniting the people and also rewarding them directly to sustain.

\begin{figure}[h!]
	\centering
	\includegraphics[width=1.0\linewidth]{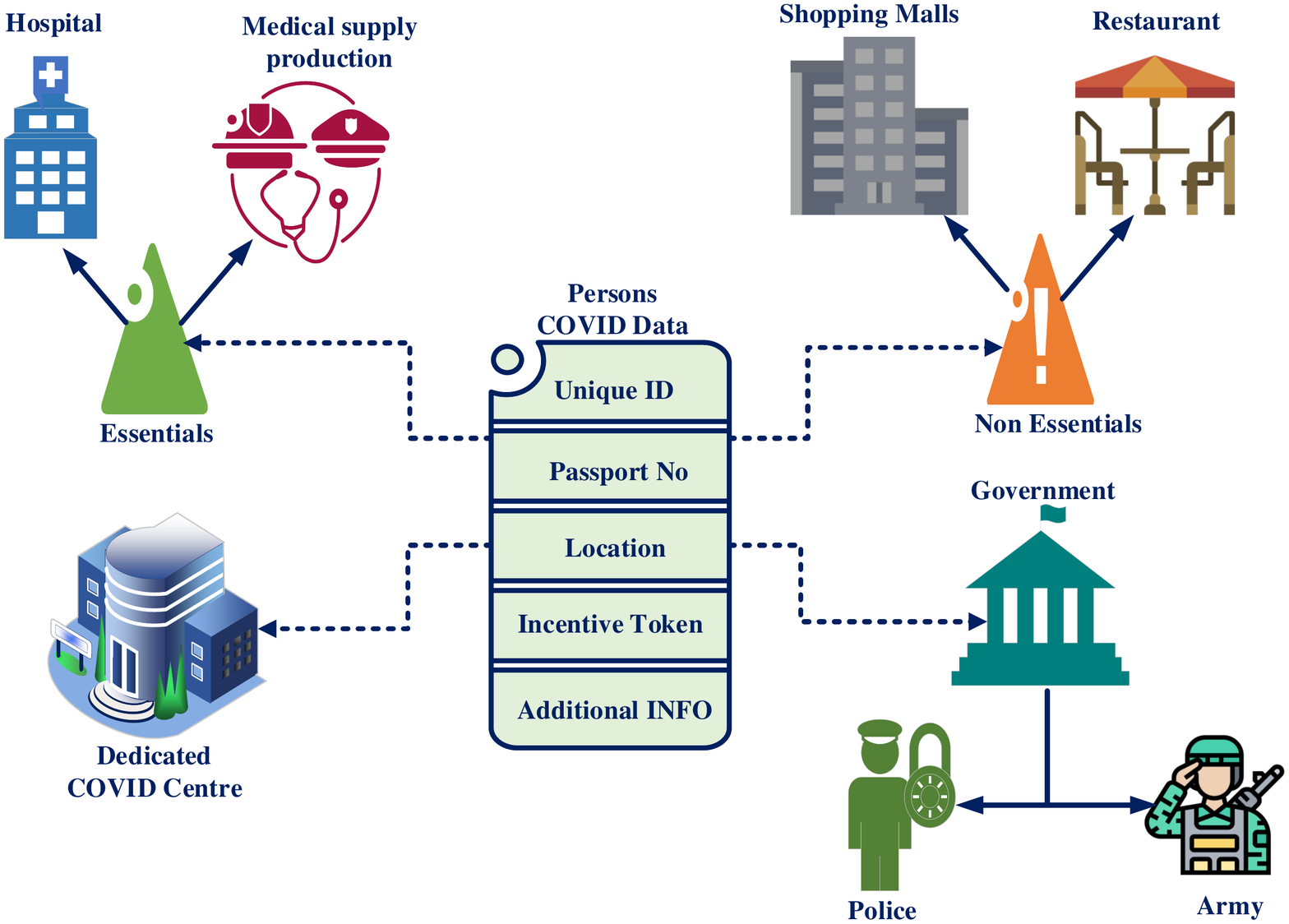}
	\caption{Visualizing interoperability of proposed solution.}
	\label{Fig:1}
\end{figure}

Life after the pandemic is still unforeseen in various regions. The sooner a country can get its economy rolling, the faster the country may grow. The fight with the pandemic is two-phased. If the economy is sacrificed to win the pandemic, the victory is bittersweet. The proposed system can be a much-needed solution to resume the declining economy. By making the COVID details easily accessible to every person, the services to healthy people may be resumed with safety measures faster than anticipated while not having major problems. The immutability and accessibility of this information are key to the revival of the economy. One of the best possible approaches to this problem is through blockchain, which can help decentralize the data while having no security concerns. Resuming services essential or non-essential, production, and opening the market are the first steps to the economic revival. The visualization of this architecture is depicted in Fig \ref{Fig:1}. 

\section{Conclusion}

The pandemic has brought uncertainty to an extent never seen before. The economic slowdown, recession, and flaws in the current health infrastructures have been brought to light during the current situation. As much as people seek the government's help, fighting a pandemic at this order of intensity is not a one mans job. To seek unity in this fight is the key to prevailing such a battle. This incentive approach takes a very different dig on the strategic battle against the pandemic by providing positive points for both the government and its people. The ease of access to COVID related data could be the much needed first step in reviving the daily activities. The strain brought by the virus to the mental and physical health of people is immeasurable, and lack of information or misinformation will only worsen the existing situation. Securing the data through blockchain and providing ease of access can help trigger the normal working of the society, which in turn can bring back the economy, providing the first steps in surviving the pandemic.
As everyone focuses on measures to prevail in the pandemic, this approach could be a nudge in the right direction with easy implementation and no drawbacks.

The proposed system prioritizes the early detection of the virus to not have negative repercussions. The work solidifies the identification of the virus spread at an early stage through the travel history of passengers. The blockchain-based system can provide a more direct self - realization approach to the current situation and become a referral backbone for the kick start of the economy while stopping the spread. Unity is a power untapped, which can give rise to newer methods and possibilities in the fight against the virus. By combining all our efforts into one problem, a solution towards that can be devised much easier. The incentive-based approach is a means to not only unite people to be on the same page but also to help people who are severely in need during these tough situations. The dreams, plans, and efforts of many have now been decapitated, and what the future holds for us has become unknown. The system can not only unite people towards the greater goal but can also help fight the pandemic in a much needed strategic manner. In the face of unpredicted danger, humanity can only rely on its unity as strength and what better way to realize that other than through technology.

\bibliographystyle{IEEEtran}
\bibliography{ref}
\end{document}